\documentclass[twoside]{article}
\usepackage{fleqn,espcrc2}

\newcommand{\AmS}{{\protect\the\textfont2
  A\kern-.1667em\lower.5ex\hbox{M}\kern-.125emS}}

\title{On M-9-branes and their dimensional reductions}

\author{Takeshi Sato\address[MCSD]{Institute for Cosmic Ray Research,
        University of Tokyo, \\
        5-1-5 Kashiwanoha, Kashiwa-shi, Chiba-ken, 277-8582, Japan}%
        }

\begin{document}

\begin{abstract}
The M-9-brane Wess-Zumino action is constructed, and by using it,
consistency of the relation of p-branes for $p \ge 8$, suggested  
on the basis of superalgebra, is discussed. 
\vspace{1pc}
\end{abstract}

\maketitle

\section{Introduction}

M-theory is a candidate for a unified theory of particle interactions
and is conjectured to be the 11-dimensional (11D) theory
which gives 5 perturbative 10-dimensional (10D) string theories 
in different kinds of limits. 
In discussing properties of these theories,
(p+1)-dimensional objects, called p-branes, 
play many crucial roles, so, it is important to clarify
what kinds of branes exist in each of the theories.
Brane scan via superalgebra is one of the methods to discuss them,
by which BPS branes possible to exist in the theories
are predicted\cite{hullalg}\cite{towalg} 
(see also \cite{towbrscan}). 
For the case $p\le 7$,
all the p-branes predicted to exist 
in M-, IIA and IIB string theories,
have corresponding solutions in each of their supergravity theories.
As for the p-branes with $p\ge 8$, however,
there is a problem, as we will explain later.
In this work we will discuss these branes,
since p-branes with $p\ge 8$ are very important in that 
M- and string theories with 16 supercharges 
are expected to be constructed by using these branes
(see ref.\cite{sptfilling} and references therein).

To be concrete,
one kind of 9-brane is suggested to exist in M-theory\cite{hullalg}, 
one kinds of 8-brane and 9-brane are predicted in
IIA, and two kinds of 9-branes are in IIB\cite{hullalg}.
The first one is called "M-9-brane", and the others are called 
or identified with D-8-brane, NS-9A-brane, D-9-brane and NS-9B-brane,
respectively, based on the consideration of their kind of charges.
Taking into account the dimensions and the duality relations of
the theories, the relation of the p-branes for 
$p\ge 8$, suggested based on superalgebra, 
is represented as 
Figure 1\cite{hullalg}.
\begin{figure}[h]
 \begin{center}
  \setlength{\unitlength}{1mm}
\begin{picture}(75,55)
\put(0,45){\makebox(20,10){D=11}}
\put(0,25){\makebox(20,10){D=10IIA}}
\put(0,5){\makebox(20,10){D=10IIB}}
\put(50,25){\dashbox(15,10){NS9A}}
\put(55,45){\vector(0,-10){10}}
\put(50,45){\framebox(15,10){M9}}
\put(50,45){\vector(-1,-1){10}}
\put(35,30){\circle{16}}
\put(30,26){\makebox(10,8){D8}}
\put(55,25){\line(0,-10){10}}
\put(35,23){\line(0,-10){6}}
\put(50,5){\dashbox(15,10){NS9B}}
\put(35,10){\circle{16}}
\put(30,5){\makebox(10,10){D9}}
\put(52,15){\makebox(12,10){T}}
\put(22,15){\makebox(12,10){T}}
\multiput(43,10)(2,0){4}{\line(1,0){1}}
\put(40,8){\makebox(12,10){S}}
\put(59,35){\makebox(10,10){Direct D.R.}}
\put(25,35){\makebox(10,10){Double D.R.}}
\end{picture}
 \end{center}
  \caption{The relation of p-branes with $p \ge 8$. 
(D.R. denotes dimensional reduction.)
}
  \label{fig:gen1}
\end{figure}
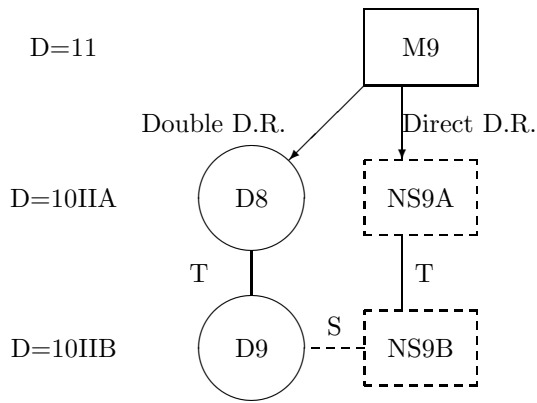

However, there is  
the following problem with the M-9-brane, or 11D origins of
the D-8-brane solution and massive IIA supergravity (SUGRA).
The BPS D-8-brane arising in the IIA string, 
actually, has a corresponding solution not in the usual IIA 
but in the massive IIA SUGRA with nonvanishing cosmological 
term\cite{pol2}\cite{pol1}\cite{berg3}. 
This is because a BPS D-8-brane in 10 dimensions is a domain wall 
with some electric charge of a RR 9-form gauge field, 
giving rise to a constant field strength, whose dual 
we denote as a mass
parameter $m$. 
This field strength contributes to
the action as a cosmological constant $-m^{2}/2$.
In other words, such domain wall
solutions cannot be constructed without cosmological term. In 11 
dimensions (11D), 
however,
no deformation to include a cosmological term
is allowed if Riemannian
geometry and covariant action are assumed\cite{des1}. 
Thus, there is no naive M-9-brane
solution in 11D, and the origin of the D-8-brane and massive IIA SUGRA
are still unclear.

There are several approaches to solve 
this problem
and one of them is "massive 11D theory"\cite{berg4}.
This is a trial theory, constructed
on the basis of the idea that {\it the
problem may imply the need to modify the framework of 11D SUGRA}.
Suppose a Killing isometry is assumed in the 11D background. Then, 
the no-go theorem is avoided and the massive 11D theory, which is
written in terms of an 11-dimensional theory at least formally,
can be defined;
it gives the 10D massive IIA SUGRA on dimensional reduction 
along the isometry direction 
(which is parametrized by the coordinate $z$), 
and gives usual 11D SUGRA in the massless limit $m \to 0$
if the dependence of the fields on $z$ is restored.
Moreover, the M-9-brane solution, i.e. the solution 
which gives a D-8-brane solution on the dimensional 
reduction along $z$, is obtained in this theory\cite{bergm9}.
We note that only the bosonic sectors have been discussed in this
massive 11D theory, though its bulk theory is called ``super''gravity.
We also note that the isometry direction is interpreted
as a compactified direction like $S^{1}$, and 
the M-9-brane is considered to
be wrapped around it\cite{bergm9}.
We follow the above idea and 
study the relation of branes within this framework.

To be concrete, we will discuss the relation from the viewpoint of
worldvolume effective action (WVEA). 
In fact, almost all of the WVEA's of the branes 
have been already obtained. However, only 
the Wess-Zumino (WZ) term of the M-9-brane $S_{M9}^{WZ}$ 
has not yet been obtained. 
So, as for the WZ terms, consistency of the relation in Figure 1
has not been established.
In other words, {\it even within this framework, 
the consistency of the
relation of the branes has not been established yet}.
The purpose of this work is to construct $S_{M9}^{WZ}$ and to examine
the consistency of the relation of the p-branes with $p \ge 8$
from the viewpoint of worldvolume effective action.

The concrete procedures are as follows:
First, we introduce a 10-form gauge potential into the theory
consistently because the M-9-brane is expected to couple to it. 
Then, we construct the M-9-brane WZ action using the 10-form
and examine the consistency of the action in certain two ways.
Finally, we investigate two kinds of dimensional reductions of
the action. 

This talk is based on work \cite{satom9,satodr}.

\section{The M-9-brane WZ action and its dimensional reductions} 
First, we briefly review the massive 11D
SUGRA\cite{berg4}.
The bosonic field content is the same 
as that of the usual (massless) 11D
SUGRA:
the metric $\hat{g}_{\mu\nu}$  and a 3-form gauge potential
$\hat{C}_{\mu\nu\rho}$. 
In this theory these fields are required to have a Killing isometry,
i.e., 
${\cal L}_{\hat{k}}\hat{g}_{\mu\nu}
={\cal L}_{\hat{k}}\hat{C}_{\mu\nu\rho}=0$
where $ {\cal L}_{\hat{k}}$ indicates a Lie derivative 
with respect to a Killing vector field $\hat{k}^{\mu}$.
(We fix the coordinates so that $\hat{k}^{\mu}=\delta^{\mu z}$.)
The infinitesimal gauge transformations of the fields are
defined as 
\begin{eqnarray}
\delta\hat{g}_{\mu\nu}=-m[\hat{\lambda}_{\mu}(i_{\hat{k}}\hat{g})_{\nu}
+\hat{\lambda}_{\nu}(i_{\hat{k}}\hat{g})_{\mu}], \nonumber\\
\delta\hat{C}_{\mu\nu\rho}=3
\partial_{[\mu}\hat{\chi}_{\nu\rho]}
-3m\hat{\lambda}_{[\mu}(i_{\hat{k}}\hat{C})_{\nu\rho]}
\label{massivegt0}
\end{eqnarray}
where 
$(i_{\hat{k}}T^{(r)}_{\mu_{1}\cdots \mu_{r-1}})\equiv\hat{k}^{\mu}
T^{(r)}_{\mu_{1}\cdots \mu_{r-1}\mu}$ for a field $T^{(r)}$.
$\hat{\chi}$ is the infinitesimal 2-form gauge parameter,
$\hat{\lambda}$ is defined as $\hat{\lambda}_{\mu}
\equiv (i_{\hat{k}}\hat{\chi})_{\mu}$, and $m$ is a constant mass
parameter.
Then, a connection for the massive gauge transformations  
should be introduced.
The new total connection takes the form 
$\hat{\Omega}_{a}^{\ bc}=\hat{\omega}_{a}^{\ bc}
+\hat{K}_{a}^{\ bc}$\footnote{
We use $a,b,\cdots$ for local Lorentz indices.}
where $\hat{\omega}_{a}^{bc}$ is a usual spin connection
and
$\hat{K}$ is given by
\begin{equation}
\hat{K}_{a}^{\ bc}=\frac{m}{2}
[\hat{k}_{a}(i_{\hat{k}}\hat{C})^{bc}
+\hat{k}_{b}(i_{\hat{k}}\hat{C})^{ac}
-\hat{k}_{c}(i_{\hat{k}}\hat{C})^{ab}].
\end{equation}
The 4-form field strength $\hat{G}^{(4)}$ of $ \hat{C}$
is defined as 
\begin{eqnarray}
\hat{G}^{(4)}_{\mu\nu\rho\sigma}=4D_{[\mu}\hat{C}_{\nu\rho\sigma]}
&\equiv & 4\partial_{[\mu}\hat{C}_{\nu\rho\sigma]}\nonumber\\
&+&3m(i_{\hat{k}}\hat{C})_{[\mu\nu}(i_{\hat{k}}\hat{C})_{\rho\sigma]}
\end{eqnarray}
where $D_{\mu}$ denotes the covariant derivative.
Then, $\hat{G}^{(4)}$ transforms covariantly as
$
\delta \hat{G}^{(4)}_{\mu\nu\rho\sigma}=4m\hat{\lambda}_{[\mu}
(i_{\hat{k}}\hat{G}^{(4)})_{\nu\rho\sigma]},
$ 
which implies that $\delta (\hat{G}^{(4)})^{2}=0$.

The action of the massive 11D SUGRA is 
\begin{eqnarray}
\hat{S}_{0}=\frac{1}{\kappa}
\int d^{11}x [\ \sqrt{|\hat{g}|}\{ \hat{R}
-\frac{1}{2\cdot 4!}(\hat{G}^{(4)})^{2}\ \ \ \ \ \ \nonumber\\
+\frac{1}{2}m^{2}|\hat{k}|^{4} \} 
+\frac{\hat{\epsilon}^{
\mu_{1}\cdots \mu_{11}}}{(144)^{2}}
\{2^{4}\partial\hat{C}\partial\hat{C}\hat{C}\ \ \ \ \ \ \ 
\nonumber\\
\ \ +18m\partial\hat{C}\hat{C}(i_{\hat{k}}\hat{C})^{2}
+\frac{3^{3}}{5}m^{2}\hat{C}(i_{\hat{k}}
\hat{C})^{4}\}_{
\mu_{1}..\mu_{11}}]
\label{11daction0}
\end{eqnarray}
where $\kappa =16\pi G_{N}^{(11)}$ and $|\hat{k}|
=\sqrt{-\hat{k}^{\mu}\hat{k}^{\nu}\hat{g}_{\mu\nu}}$.
This action is invariant (up to total derivative)
under (\ref{massivegt0}), and
its dimensional reduction 
along $z$
gives the bosonic part of 10D massive IIA SUGRA action.

Now, let us introduce a 10-form gauge potential $\hat{A}^{(10)}$.
Following the case of the 9-form potential in 10D IIA 
theory\cite{berg3},
we promote
the mass parameter $m$ to a scalar field $\hat{M}(x)$,
and add the term 
\begin{eqnarray}
\frac{1}{\kappa}\int d^{11}x 
\frac{11}{11!}\hat{\epsilon}^{\mu_{1}\cdots\mu_{11}}\hat{M}(x)
\partial_{[\mu_{1}} \hat{A}^{(10)}_{\mu_{2}\cdots\mu_{11}]}.  
\end{eqnarray}
to the action $\hat{S}_{0}$
to introduce $\hat{A}^{(10)}$ as a Lagrange multiplier for 
$\hat{M}(x)=m$.
We note that $\hat{A}^{(10)}$ also satisfies $ {\cal
L}_{\hat{k}}\hat{A}^{(10)}=0$, which means that
$\hat{A}^{(10)}$ with no $z$ index does not appear in this theory.
Then, the action is invariant under (\ref{massivegt0})
if the massive gauge transformation of $\hat{A}^{(10)}$
is defined as
\begin{eqnarray}
\delta (i_{\hat{k}}\hat{A}^{(10)})_{\mu_{1}\cdots\mu_{9}}
=-\sqrt{|\hat{g}|}\hat{\epsilon}_{\mu_{1}\cdots\mu_{9} z}
\ \ \ \ \ \ \ \ \ \ \ \ 
\nonumber\\
\ \ \ \cdot[-\hat{g}^{\mu\mu'}\hat{g}^{\nu\nu'}
(2\partial_{[\mu'} \hat{k}_{\nu']}
-\hat{M}|\hat{k}|^{2}(i_{\hat{k}}\hat{C})_{\mu'\nu'})
\hat{\lambda}_{\nu}\nonumber\\
\ \ \ \ \ \ \ +\frac{1}{2}\hat{G}^{(4)\mu\nu\rho\sigma}
(i_{\hat{k}}\hat{C})_{\nu\rho}\hat{\lambda}_{\sigma}]\nonumber\\
\ \ \ \ \ 
-\frac{9!}{48}[\partial \hat{C}(i_{\hat{k}}\hat{C})^{2}\hat{\lambda}
+\frac{\hat{M}}{4}(i_{\hat{k}}\hat{C})^{4}
\hat{\lambda}]_{\mu_{1}\cdots\mu_{9}}.\label{10formtr2}
\label{10formtr1} 
\end{eqnarray}

Now, we discuss
the construction of $S_{M9}^{WZ}$ 
using $\hat{A}^{(10)}$.
However, the gauge invariant WZ action
cannot be constructed straightforwardly.
The reason is as follows:
Since a M-9-brane couples to $\hat{A}^{(10)}$,
$S_{M9}^{WZ}$ contains the term
\begin{eqnarray}
S_{M9}^{WZ}|_{{\rm 9form}}=
\frac{T_{M9}}{9!}\int d^{9}\xi\  \epsilon^{i_{1}.. i_{9}}
\ \ \ \ \ \ \ \ \ \ \ \ \ \ \ 
\nonumber\\
\ \ \ \ \ \ \ \ \ \ \ \ \ \times
\partial_{i_{1}}X^{\hat{\mu}_{1}}..\partial_{i_{9}}X^{\mu_{9}}
(i_{\hat{k}}\hat{A}^{(10)})_{\mu_{1}..\mu_{9}}\ \ \ \label{d8brwz}
\end{eqnarray}
where $\xi^{i}$ ($i=0,..,8$) are worldvolume coordinates
of the brane and $\hat{X}^{\mu} \ \ (\mu=0,..,9)$ are embedding coordinates.
Suppose we consider the massive gauge transformation of (\ref{d8brwz}).
Then, we can see that 
the contribution of the first bracket of 
the r.h.s. of (\ref{10formtr2})
to the variation of (\ref{d8brwz})
cannot be cancelled
even if any other terms are added to (\ref{d8brwz}).
This is because 
the contribution of the bracket cannot be represented by any
products of forms due to the extra $\hat{\epsilon}$
but that all the terms of $S_{M9}^{WZ}$
should be represented by some products of forms.

Our idea to resolve this problem is as follows:
Since the main obstruction is the existence of 
the extra $\hat{\epsilon}$ in (\ref{10formtr2}),
let us suppose one rewrite the first bracket of (\ref{10formtr2})
by using the ``dual fields'' of $\hat{k}_{\hat{\mu}}$ and $\hat{C}$
through duality relations.
Then, the extra $\hat{\epsilon}$ is cancelled and
the first bracket can be expressed 
as a sum of exterior products of forms.
Thus, it is expected that one can construct
a gauge invariant WZ action.
This idea is successful,
which we show in the following.

The dual field of the 3-form $\hat{C}$
is the 6-form $\hat{C}^{(6)}$ 
whose massive gauge transformation, field strength and
the duality relation are\cite{berg4}\footnote{We
concentrate our discussions on the gauge 
transformations with respect to
$\hat{\chi}$ and $\hat{\lambda}$.} 
\begin{eqnarray}
\delta \hat{C}^{(6)}_{\mu_{1}\cdots\mu_{6}}&=&30
\partial_{[\mu_{1}}\hat{\chi}_{\mu_{2}\mu_{3}}
\hat{C}_{\mu_{4}\mu_{5}\mu_{6}]}\nonumber\\  
&+&6\hat{M}\hat\lambda_{[\mu_{1}}
(i_{\hat{k}}\hat{C}^{(6)})_{\mu_{2}\cdots\mu_{6}]}\\
\hat{G}^{(7)}_{\mu_{1}\cdots\mu_{7}}&=&7[
\partial\hat{C}^{(6)}
-3\hat{M}(i_{\hat{k}}\hat{C})(i_{\hat{k}}\hat{C}^{(6)})\nonumber\\
&+&10\hat{C}\partial \hat{C}
+5\hat{M}C(i_{\hat{k}}\hat{C})^{2}\nonumber\\
& &+\frac{\hat{M}}{7}(i_{\hat{k}}\hat{N}^{(8)})]_{\mu_{1}\cdots\mu_{7}}\\
\hat{G}^{(4)\mu_{1}\cdots\mu_{4}}&=&\frac{\epsilon^{\mu_{1}\cdots
\mu_{11}}}{7!\sqrt{|\hat{g}|}}\hat{G}^{(7)}_{\mu_{5}\cdots
\mu_{11}}\label{11ddual1}.
\end{eqnarray}
$\hat{N}^{(8)}$ is 
the dual field of the Killing vector also introduced in
ref.\cite{berg4}, whose gauge transformation is suggested such as
\begin{eqnarray}
\delta \hat{N}^{(8)}_{\mu_{1}\cdots\mu_{8}}&=&\{
\frac{8!}{3\cdot 4!}\partial \hat{\chi}\hat{C}(i_{\hat{k}}\hat{C})
\nonumber\\
& &+8\hat{M}\hat{\lambda}(i_{\hat{k}}\hat{N}^{(8)})\}
_{\mu_{1}\cdots\mu_{8}}.
\end{eqnarray}
In this paper 
we regard 
$\hat{k}_{\mu}\equiv (i_{\hat{k}}\hat{g})_{\mu}$ as a ``vector
gauge field'', 
and consider the ``field strength'' of it,
as one does for 
$(i_{\hat{k}}\hat{C})$.
Then, if we define $\hat{G}^{(2)}$
as
\begin{eqnarray}
\hat{G}^{(2)}_{\mu\nu}\equiv 2\partial_{[\mu}\hat{k}_{\nu]}
-\hat{M}|\hat{k}|^{2}(i_{\hat{k}}\hat{C})_{\mu\nu},
\end{eqnarray}
$\hat{G}^{(2)}$ is shown to 
transform covariantly under (\ref{massivegt0}).
So, $\hat{G}^{(2)}$, in fact 
arising in the first term of (\ref{10formtr2}),
can be interpreted as the field strength of 
$\hat{k}_{\mu}$.
On the other hand,
the field strength $\hat{G}^{(9)}$ 
of the full 8-form $\hat{N}^{(8)}$
is difficult to construct.
However, in order to rewrite the first term through the duality
relation between $\hat{G}^{(9)}$ and $\hat{G}^{(2)}$,
it is sufficient to know
the field strength of $(i_{\hat{k}}\hat{N}^{(8)})$. 
This is because $\hat{G}^{(2)}$
in the first term of (\ref{10formtr1})
vanishes if
any of the indices of $\hat{G}^{(2)}$ takes $z$,
implying that one of the indices of $\hat{G}^{(9)}$ certainly 
takes $z$. 
Thus,
only the field strength of $(i_{\hat{k}}\hat{N}^{(8)})$ is needed, 
and it can be defined as
\begin{eqnarray}
(i_{\hat{k}}\hat{G}^{(9)})_{\mu_{1}\cdots\mu_{8}}\equiv
8\{ \partial (i_{\hat{k}}\hat{N}^{(8)})
\ \ \ \ \ \ \ \ \ \ \ \ \ \nonumber\\
\ \ \ \ \ \ 
+21(i_{\hat{k}}\hat{C}^{(6)})\partial(i_{\hat{k}}\hat{C})
+35C\partial(i_{\hat{k}}\hat{C})(i_{\hat{k}}\hat{C})\nonumber\\
\ \ \ \ \ \ \ +35\partial C(i_{\hat{k}}\hat{C})^{2}
+\frac{105}{8}\hat{M}(i_{\hat{k}}\hat{C})^{4}
\}_{[\mu_{1}\cdots\mu_{8}]}.\label{n8fs}
\end{eqnarray}
We note that $(i_{\hat{k}}\hat{G}^{(9)})$ 
is invariant under (\ref{massivegt0}), which means that this
definition is consistent.
Then, we assume the duality relation: 
\begin{eqnarray}
\hat{G}^{(2)\mu_{1}\mu_{2}}=\frac{\epsilon^{\mu_{1}\cdots
\mu_{10}z}}{9!\sqrt{|\hat{g}|}}
(i_{\hat{k}}\hat{G}^{(9)})_{\mu_{3}\cdots\mu_{10}}\label{11ddual2}.
\end{eqnarray}
It gives
one of the 10D IIA duality relations in ref.\cite{DWZ}
on dimensional reduction in $z$, which means that (\ref{11ddual2}) 
is consistent. 

Since all the preparations have been done,
let us substitute the relation
(\ref{11ddual1}) and (\ref{11ddual2}) for
(\ref{10formtr2})
to have the rewritten expression
of the massive gauge transformation of $\hat{A}^{(10)}$:
\begin{eqnarray}
\delta (i_{\hat{k}}\hat{A}^{(10)})_{\mu_{1}\cdots\mu_{9}}
&=&-9![\ 
\frac{1}{7!}\partial(i_{\hat{k}}\hat{N}^{(8)})\hat{\lambda}
\nonumber\\
&-&\frac{1}{2\cdot 5!}\partial\{(i_{\hat{k}}\hat{C}^{(6)})
(i_{\hat{k}}\hat{C})\}\hat{\lambda}\nonumber\\
&+&\frac{1}{6\cdot 4!}\partial\{\hat{C}(i_{\hat{k}}\hat{C})^{2}\}
\hat{\lambda}\nonumber\\
&-&\frac{\hat{M}}{2^{4}\cdot 4!}
(i_{\hat{k}}\hat{C})^{4}\hat{\lambda}\ 
]_{[\mu_{1}\cdots\mu_{9}]}.
\end{eqnarray}
By using this expression,
the gauge invariant WZ action of the M-9-brane can be constructed indeed.
Before constructing it, we give
the rewritten field equation of $\hat{M}(x)$:
\begin{eqnarray}
-\hat{M}|\hat{k}|^{4}
=\frac{10\hat{\epsilon}^{\mu_{1}..\mu_{10}z}}{10!\sqrt{|\hat{g}|}}
\{\partial_{\mu_{1}}
(i_{\hat{k}}\hat{A}^{(10)})_{\mu_{2}..\mu_{10}}\nonumber\\
-\frac{9!}{8\cdot 6!}(i_{\hat{k}}\hat{G}^{(7)})(i_{\hat{k}}\hat{C})^{2}
+\frac{9!}{2\cdot 8!}(i_{\hat{k}}\hat{G}^{(9)})(i_{\hat{k}}\hat{C})
\nonumber\\
+\frac{9!}{288}
\partial\hat{C}(i_{\hat{k}}\hat{C})^{3}
+\frac{9\cdot 9!}{5760}\hat{M}(i_{\hat{k}}\hat{C})^{5}
\}_{\mu_{1}..\mu_{10}}.\label{fs11form}
\end{eqnarray}
Since the r.h.s. of (\ref{fs11form}) 
is shown to be gauge invariant,
it can be interpreted as the gauge invariant field strength of 
the 10-form (multiplied by 1/10!).
Thus, we can conclude that the 10-form $\hat{A}^{(10)}$
is introduced consistently.
Moreover,
we define 
a new 10-form $\hat{C}^{(10)}$
which agrees with 10D IIA 9-form $C^{(9)}$
on dimensional reduction along z:
\begin{eqnarray}
(i_{\hat{k}}\hat{C}^{(10)})_{\mu_{1}\cdots\mu_{9}}
\equiv (i_{\hat{k}}\hat{A}^{(10)})_{\mu_{1}\cdots\mu_{9}}
\ \ \ \ \ \ \ \ \ \ \ \ \ 
\nonumber\\
+[\frac{9!}{2\cdot
7!}(i_{\hat{k}}\hat{N}^{(8)})(i_{\hat{k}}\hat{C})
-\frac{9!}{2^{3}5!}(i_{\hat{k}}\hat{C}^{(6)})
(i_{\hat{k}}\hat{C})^{2}\nonumber\\
+\frac{9!}{2^{4}(3!)^{2}}
\hat{C}(i_{\hat{k}}\hat{C})^{3}]_{[\mu_{1}..\mu_{9}]}
\end{eqnarray}
Then, the gauge transformation of $\hat{C}^{(10)}$ takes the simple form:
\begin{eqnarray}
\delta (i_{\hat{k}}\hat{C}^{(10)})_{\mu_{1}\cdots\mu_{9}}
&=&-945\{-4\partial\hat{\chi}(i_{\hat{k}}\hat{C})^{3}\nonumber\\
& &+ \hat{M}(i_{\hat{k}}\hat{C})^{4}\hat{\lambda}\ 
\}_{[\mu_{1}\cdots\mu_{9}]}.
\end{eqnarray}
For convenience, we use $\hat{C}^{(10)}$
to construct $S^{WZ}_{{\rm M9}}$.

Now, we construct the M-9-brane WZ action
as that of the gauged $\sigma$-model,
in which the translation along $\hat{k}$ is 
gauged\cite{kaluzakleinm}\cite{berg4}\cite{loz1}.
In this approach the M-9-brane
wrapped around the compact isometry direction is 
described\cite{bergm9}.
So, denoting its worldvolume coordinates by $\xi^{i} \  (i=0,1,..,8)$
and their embeddings by $X^{\mu}(\xi) (\mu =0,1,..,9,z)$,
the worldvolume gauge
transformation is given by
$\delta_{\eta} X^{\mu}=\eta(\xi)\hat{k}^{\mu}$
where 
$\eta(\xi)$ is a scalar gauge parameter.
In order to make the brane action invariant
under the transformation,
the derivative of $X^{\mu}$ with respect to $\xi^{i}$ is 
replaced by
the covariant derivative
$D_{i}X^{\mu}=\partial_{i}X^{\mu}
-\hat{A}_{i}\hat{k}^{\mu} $
with the gauge field $\hat{A}_{i}=-|\hat{k}|^{-2}
\partial_{i}\hat{X}^{\hat{\nu}}\hat{k}_{\hat{\nu}}$\cite{loz1}.
Then, we obtain
the M-9-brane WZ action 
only on the basis of the gauge invariance, as 
\begin{eqnarray}
S_{M9}^{WZ}=T_{{\rm M9}}
\int d^{9}\xi \epsilon^{i_{1}\cdots i_{9}} 
[\frac{1}{9!}\widetilde{(i_{\hat{k}}\hat{C}^{(10)})}_{i_{1}\cdots
i_{9}}
\nonumber\\
+\frac{1}{2\cdot 7!}
\widetilde{(i_{\hat{k}}\hat{N}^{(8)})}_{i_{1}\cdots i_{7}}
\hat{{\cal K}}^{(2)}_{i_{8}i_{9}} 
\nonumber\\
+\frac{1}{2^{3}\cdot 5!}\widetilde{(i_{\hat{k}}\hat{C}^{(6)})}_{
i_{1}\cdots i_{5}}
(\hat{{\cal K}}^{(2)})^{2}_{{i_{6}\cdots i_{9}}} \nonumber\\
+\frac{1}{2\cdot (3!)^{2}}\widetilde{\hat{C}}_{i_{1} i_{2} i_{3}}
\{(\partial \hat{b})^{2} 
-\frac{1}{4}\widetilde{(i_{\hat{k}}\hat{C})}
\partial \hat{b} \nonumber\\
+\frac{1}{8}\widetilde{(i_{\hat{k}}\hat{C})}^{2}\}_{i_{4}\cdots i_{7}}
\hat{{\cal K}}^{(2)}_{i_{8} i_{9}}\nonumber\\
+\frac{1}{2\cdot 4!}\hat{A}_{i_{1}}
\{ (\partial \hat{b})^{3} 
+\frac{1}{2}(\partial \hat{b})^{2}\widetilde{(i_{\hat{k}}\hat{C})}
\nonumber\\
+\frac{1}{4}(\partial \hat{b})\widetilde{(i_{\hat{k}}\hat{C})}^{2}
+\frac{1}{8}\widetilde{(i_{\hat{k}}\hat{C})}^{3}
\}_{i_{2}\cdots i_{7}}
(\hat{{\cal K}}^{(2)})_{i_{8} i_{9}}\nonumber\\
+\frac{m}{5!}\hat{b}_{i_{1}}(\partial \hat{b})^{4}_{i_{2}\cdots
i_{9}}]
\label{m9action}
\end{eqnarray}
where $\widetilde{\hat{S}}_{i_{1}\cdots i_{r}}\equiv 
\hat{S}_{\mu_{1}\cdots\mu_{r}}
D_{i_{1}}X^{\mu_{1}}\cdots D_{i_{r}}X^{\mu_{r}}$
for a target-space field $\hat{S}_{\mu_{1}\cdots\mu_{r}}$.
$\hat{b}_{i}$ 
describes the flux 
of an M-2-brane wrapped around the isometry direction,
whose massive gauge transformation is determined by the 
requirement of the invariance of its modified field strength
$\hat{{\cal K}}^{(2)}_{ij}=2\partial_{[i}\hat{b}_{j]}-
\partial_{i}X^{\mu}\partial_{j}X^{\nu}
(i_{\hat{k}}\hat{C})_{\mu\nu}$ (i.e. $\delta\hat{b}_{i}
=\hat{\lambda}_{i}$).

Then, we check the consistency of the M-9-brane action in two ways; 
(the kinetic term of the M-9-brane has been 
given in ref.\cite{eyras1}.)
first, we can improve
the M-9-brane solution in ref.\cite{bergm9}
so that $\hat{A}^{(10)} \ne 0$.
Then, the M-9-brane worldvolume action must be the source of the
solution. We can show that this is true\cite{satom9}.
Second, when there are
two M-9-branes parallel to each other with a certain 
orientation, no force exists between them,
so, the potential energy of a test M-9-brane 
parallel to a background M-9-brane must vanish.
Using the obtained M-9-brane action and the improved
M-9-brane solution,
we can show that this is also true\cite{satom9}.
Thus, we can say that the obtained M-9-brane action is consistent.

Finally, we present the result of dimensional reductions of
$S_{M9}^{WZ}$ briefly.
First, if we consider the dimensional reduction along the isometry
direction, it is shown to give the D-8-brane  WZ action.
Second, if we consider the dimensional reduction along the only
transverse direction, $S_{M9}^{WZ}$ is shown to give the NS-9A-brane  
WZ action.
(In fact, in this case, we need to know ``undiscussed'' truncation
conditions caused by modding out the system by an certain $Z_{2}$
symmetry, but we can infer them by using the duality relations
(\ref{11ddual1}) and (\ref{11ddual2}).)
Thus, the relation of p-branes with $p \ge 8$, based on the
superalgebra, is consistent 
from the viewpoint of their WVEAs.

\section{Summary and discussion} 
The results of this work is summarized as follows:
The M-9-brane Wess-Zumino
action, the only unconstructed (bosonic part of) brane action,
has been obtained, based only on the gauge invariance.
The essential point in constructing it 
is our appropriate choice of fields representing the same degrees of
freedom.
Its consistency has been confirmed in two ways.
In addition, upon two kinds of dimensional reductions, the
Wess-Zumino action of the M-9-brane has been shown to give 
those of the D-8-brane and the NS-9A brane, respectively.
Therefore, we conclude that within the framework of  
 massive 11D theory, the relation of p-branes with 
$p \ge 8$, suggested on the basis of
superalgebra, is consistent 
from the viewpoint of their worldvolume effective actions.

In this theory, however, 
the implication of the existence of the isometry direction 
is still unclear, so some other modification of the framework might be 
needed.

Finally, we would like to note that
there is third possibility of dimensional reduction of the M-9-brane;
the dimensional reduction along the worldvolume direction but not the
isometry one. There are some arguments on how to interpret this 
possibility, and ours is that the obtained 8-brane is essentially
the same as the usual D-8-brane
except that it arises in another massive extension of the 
10-dimensional IIA theory with an isometry direction.
Since we do not have enough space to discuss it here,
in detail, please see ref.\cite{satodr} and references therein.
\vspace{0.5pc}

\noindent
{\large \bf Acknowledgment}

I would like to thank Taro Tani, Tunehide Kuroki and Shinya Tamura 
for fruitful discussions and encouragement in completing 
the work in refs.\cite{satom9}\cite{satodr}.
I am grateful to Professor Eric Bergshoeff for useful comments
and Yolanda Lozano for useful comments on the work\cite{satodr} 
via e-mail.
I am especially grateful to Professor Dmitri Sorokin,
Professor Alexei Nurmagambetov and all the other staffs and students
supporting the conference, 
for inviting me to the conference and taking much care of me and my
wife very kindly before and during the conference.
I am also grateful to Yoshida Foundation for Science and Technology
for partial financial support.

\end{document}